\documentstyle[11pt,paspconf,epsf]{article}

\begin{document}

\title{Near-Infrared Imaging of a Group or Cluster of Galaxies at a Redshift 
of 2.39}

\author{Ian Waddington}
\affil{Institute for Astronomy, University of Edinburgh, Blackford Hill, 
Edinburgh, EH9 3HJ, U.K.}

\begin{abstract}
At $z=2.39$, the cluster around 53W002 is one of the most distant
groups or clusters of galaxies known to date.  At this redshift the
4000\AA-break falls between the $J$ and $H$ bands, thus our infrared 
observations are designed to identify cluster members by a red $J-H$
colour.  Out of the 42 objects we have detected in the field, we find
ten galaxies with $J-H>1.0$ and $K>18.8$, redder and fainter than the
radio galaxy, and consistent with the presence of a 4000\AA-break at
the cluster redshift.  Two of these reddest galaxies have been
confirmed spectroscopically.  The colours, sizes and location of these
infrared-selected galaxies suggest a cluster much more similar to
those nearby than revealed by {\it Hubble Space Telescope\/}
observations alone.
\end{abstract}


\section{Introduction} 

An important recent breakthrough in the search for high-redshift
galaxies has been the technique of Lyman-limit imaging, pioneered by
Steidel, Pettini \& Hamilton~(1995).  They discovered a substantial
population of star-forming galaxies at $3.0<z<3.5$, selected using a
custom set of broadband filters designed to identify the presence of
the 912\AA\ Lyman break at these redshifts.  The successful
confirmation of their redshifts via deep Keck spectroscopy (Steidel 
et~al.~1996a) has led to many more applications of this technique (for
example, Steidel et~al.~1996b; Stevens, Lacy \& Rawlings~1998).
However, this selection method is strongly biased towards blue
star-forming objects, and a similar bias is also present in
narrow-band searches designed to identify high-redshift objects via
strong emission lines (for example, Hu, McMahon \& Egami~1996; Malkan,
Teplitz \& McLean~1996).  This is unfortunate since the passively
evolving stellar population of a galaxy can be masked with relative
ease by even a small starburst, and it is the reddest, oldest,
most-passive systems at any redshift which are of greatest importance
for constraining the epoch of elliptical galaxy formation and indeed
the age of the Universe.

\begin{figure}[tb]
\plotone{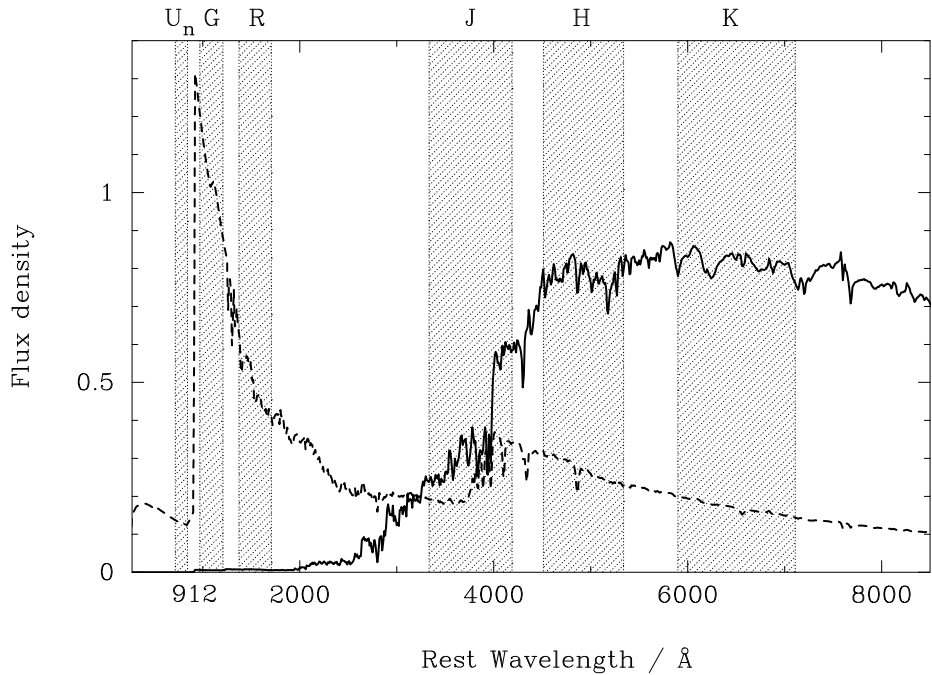}
\caption{Comparison of the Lyman-limit (912\AA) and 4000\AA-break selection
techniques.  The two spectra are Guiderdoni \& Rocca-Volmerange~(1987)
models of ages 1 Gyr (dashed line) and 9 Gyr (solid line).  Shading
denotes the filter bandpasses at redshifts of 3.25 for $U_n$, $G$, $R$, 
and at 2.39 for $J$, $H$, $K$.}
\label{select} 
\end{figure}

In order to find these passively evolving galaxies one needs a
selection technique that is sensitive to an old stellar population
rather than a young one.  Such a diagnostic is the 4000\AA-break ---
the step in a galaxy's spectrum due to Balmer continuum absorption,
calcium H and K lines and, at a slightly larger wavelength, the iron
G-band.  The strength of the break is indicative of the age of the
stellar population --- an evolved galaxy will show a large break,
whereas a young/star-forming galaxy will have a much weaker break.
This should be contrasted with the Lyman continuum break at 912\AA\
used by Steidel et~al., which is strong for a young galaxy but very
weak for an old one (figure~\ref{select}).  Thus these two breaks
provide complementary selection methods, one biased towards blue
active galaxies and one biased towards red passively evolving
galaxies.

It is not necessary to produce customized filters (such as were used
by Steidel et~al.) to measure the strength of the
4000\AA-break, so long as one is prepared to be limited to the
redshifts for which the break falls between standard filters; for
example, $z \sim 1.6$ for $I$ \& $J$, $z \sim 2.7$ for $J$ \& $H$, and
$z \sim 3.8$ for $H$ \& $K$.  Figure~\ref{select} illustrates how the
$J$ and $H$ filters approximately bridge the break for an
evolved galaxy at $z=2.39$.  This led us to the idea of using
near-infrared imaging to identify galaxies with a strong break on the
basis of their infrared colours.

The target field chosen for this project was that around the radio
galaxy 53W002, which has a redshift of 2.39.  {\it Hubble Space
Telescope\/} and ground-based searches for emission-line objects
surrounding this galaxy revealed a total of eighteen potential cluster
members.  Spectroscopic redshifts of $z\approx 2.4$ have now been
measured for several of these candidates (Pascarelle et~al.~1996a,b).
The compact size of these sources and their faint luminosities have
suggested that they may be the subgalactic-sized progenitors of
massive ellipticals or spirals.  Once again though, the selection
technique used to find these sources is biased towards those galaxies
with detectable Ly$\alpha$ emission, i.e.~active or star-forming
galaxies, and is insensitive to any passively evolving cluster
members.  The complementary search for such passive galaxies, reported
here, will thus allow us to investigate the full range of
star-formation activity in this cluster.

\section{Observations}

The field was observed at the 3.8m UK Infrared Telescope, Mauna Kea,
Hawaii on 26--27 May 1995, using the near-infrared imaging camera
IRCAM3, in reasonable weather and with seeing of 0.8--1.2 arcsec.  A
standard jittering procedure was used to obtain a median-filtered sky
flat-field simultaneously with the data.  Total exposure times of
either 54 or 81 minutes were obtained in $J$, $H$ and $K$ across a
field of $\sim$100$\times$100 arcsec$^{2}$.

Magnitudes for the sources were measured in a 3.5-arcsec diameter
aperture.  The reliability of the photometry was tested by two of us
calculating magnitudes independently for a selection of faint sources
across the field, and the results were found to be in good agreement.
Our magnitudes for the radio galaxy ($J=20.70\pm 0.19$, $H=19.77\pm
0.28$ \& $K=18.81\pm 0.12$) are also consistent with those of previous
observations ($J=20.70\pm 0.28$, $H=19.84\pm 0.29$ \& $K=18.61\pm
0.17$; Windhorst et~al.~1994).

\section{Results \& Discussion} 

\begin{figure}[tb]
\plotone{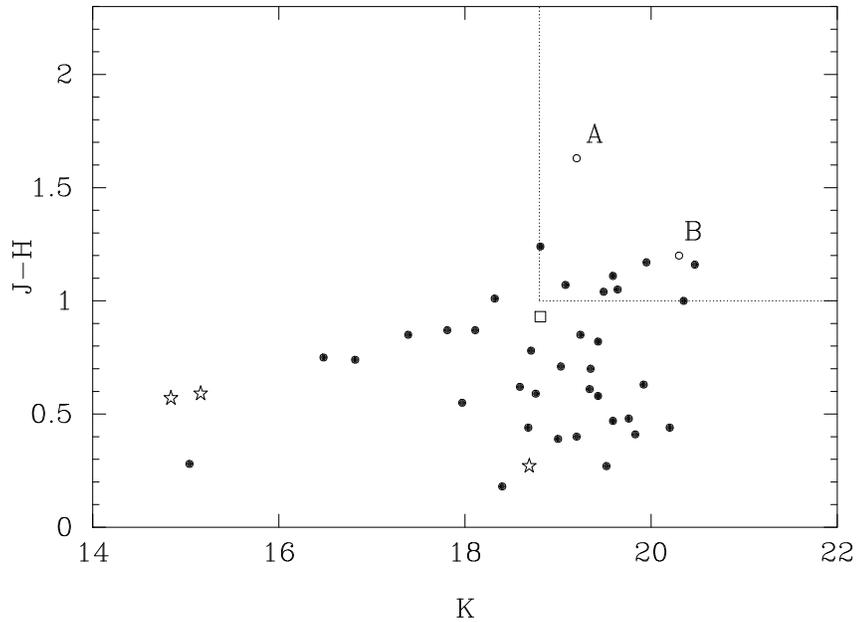}
\caption{Colour--magnitude diagram for the 42 sources in the field detected 
          at $K$.  The radio galaxy is denoted by a box; spectroscopically 
          confirmed members of the cluster by a hollow circle (A \& B); and 
          a star denotes a source which appears stellar on the {\it HST\/} 
	  image of Pascarelle et~al.~(1996b).  Dotted lines show the criteria 
	  used to select potential cluster members ($J-H>1.0$, $K>18.8$).}
\label{colmag}
\end{figure}

We detect a total of 42 objects in the $K$-band, of which 36 are above
our 80\% completeness limit.  Of these 42 sources, we have isolated a
sample of 10 galaxies with $J-H>1.0$ and $K>18.8$, that are redder and
fainter than the radio galaxy, and consistent with the presence of a
4000\AA-break at $z=2.39$ (figure~\ref{colmag}).  Our selection
criteria are based on the following: (i) radio galaxies can have a
significant contribution to their blue flux arising from the embedded
active nucleus, so we would expect any passively evolving galaxy at
the same redshift, in the absence of an AGN, to possess a stronger
4000\AA-break and hence have a redder colour; (ii) the radio galaxy
has above-average optical luminosity (${\rm L}\approx {\rm L}^*$,
Windhorst et~al.~1991), so we expect any companion galaxies to be
comparable to, or fainter than, 53W002 in $K$ (rest-frame optical).  The
selection rules are illustrated on the colour--magnitude diagram as
dotted lines.

Below we will compare the colours of our red cluster candidates with
those predicted by spectro-photometric models of elliptical galaxies,
but first we consider model-independent evidence that these ten
sources are indeed part of the cluster.  Most excitingly, the reddest
and third reddest galaxies in our sample (objects A and B; denoted
`19' and `18' in Pascarelle et~al.~1996b) are the only two sources in
the cluster, that are detected at $K$, which have been shown via
optical spectroscopy to lie at the same redshift as the radio galaxy.
The other spectroscopically confirmed members were either not
detected in the $K$-band or lay outside the infrared field.  The
remaining eight of our ten red galaxies have not been identified as
Ly$\alpha$ emitters, and are good candidates for passively evolving
elliptical galaxies at $z=2.39$, with significant breaks and, 
at most, weak Ly$\alpha$ emission.  These extremely encouraging
results show that (i) near-infrared colours can indeed be used to
successfully identify high-redshift galaxies on the basis of the
strength of the 4000\AA-break, and (ii) at least some galaxies
possess a significant 4000\AA-break at large lookback times, where
most galaxies studied to date have been found to be
relatively blue.

\begin{figure}[tb]
\plotone{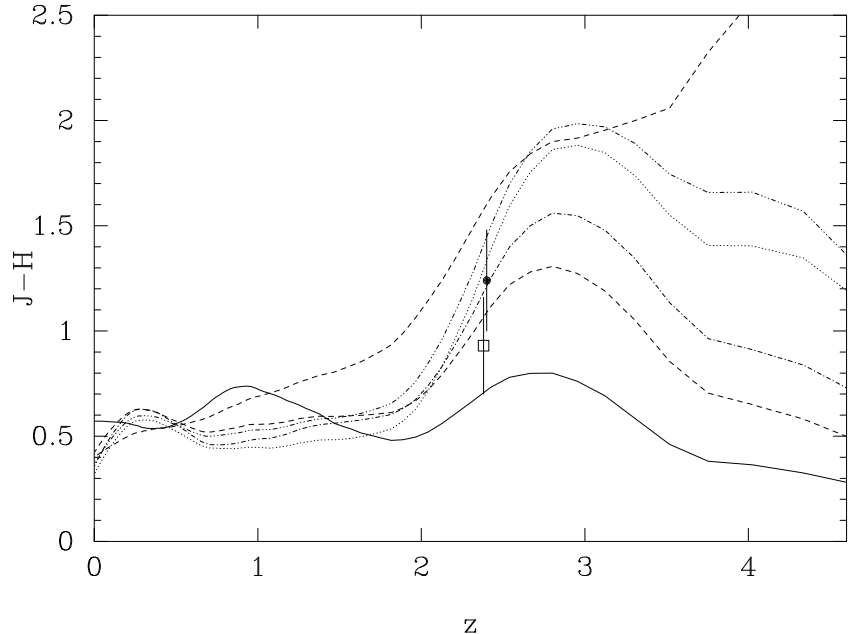}
\caption{The predicted dependence of $J-H$ colour on redshift for a 1 Gyr 
          burst model of galaxy evolution at ages of 0.1 Gyr (solid line), 
          0.6 Gyr (dashed line), 1.1 Gyr (dot-dash line), 1.6 Gyr (dotted 
          line), 2.1 Gyr (dot-dot-dot-dash line) and 6.1 Gyr (upper 
	  dashed line), compared with the observed $J-H$ colour of the 
	  radio galaxy (box) and the second reddest companion galaxy at 
	  $z=2.39$ (circle).}
\label{jhmodels}
\end{figure}

It is particularly interesting to compare these sources with those in
the 2139$-$4434B1 group of Francis et~al.~(1997; also, this volume).
Three galaxies (B1, B2, B4) at a redshift of 2.38 have been discovered
via narrow-band Ly$\alpha$ imaging, and confirmed spectroscopically.
B1 has the same infrared magnitudes as our object A, and (after making
a correction for the AGN contribution in A) the two galaxies have
comparable optical magnitudes.  The colours of all three galaxies
($J-K=2.1$--2.8) are also comparable to the four reddest of our
sources ($J-K=1.9$--2.6), although the mean colour of our sources is a
little less at $J-K=1.7$.

In figure~\ref{jhmodels} we present results of our modelling of the
evolution of the second reddest of our cluster members, using an
updated version of the spectral evolution models of Guiderdoni \&
Rocca-Volmerange~(1987).  Note that we use the second reddest source,
as the reddest one, object A, is an AGN with $>$75\% of its optical
flux in an unresolved point source and 20--30\% of its $K$ flux due to
H$\alpha$ emission --- it's red colour is unlikely to be due solely to an
old stellar population.  One of the simplest scenarios to consider is
to assume that all the stars formed in a single burst of
star-formation of 1 Gyr duration and then evolved passively.  The
strength of the 4000\AA-break is then determined by the evolution of
the main sequence turn-off mass.  The $J-H$ colour of such a galaxy as
a function of redshift is shown for galaxies of different ages
(i.e.~time after the cessation of star-formation) in
figure~\ref{jhmodels}.

It can be seen that at the cluster redshift, 53W002 has the colours of
a 0.4 Gyr-old galaxy, in agreement with the age previously inferred
from more detailed spectroscopic information (Windhorst et~al.~1991).
The second reddest source in our field is predicted to have an age of
$1.1_{-0.6}^{+1.0}$ Gyr.  For comparison, an Einstein-de~Sitter
universe is only 1.4$h_{75}^{-1}$ Gyr old at this redshift.

An analysis of the infrared surface brightness profiles of
the red galaxies shows that most of the sources are clearly
resolved, with the notable exception of object A.  Fitting the mean
profile with a seeing-convolved de Vaucouleur's model gives a scale
length of $13\pm10 h_{75}^{-1}$ kpc, significantly larger than the
0.5--1.0 kpc sizes of the Ly$\alpha$-selected cluster candidates.  All
the infrared-selected galaxies and most of the optically-selected ones
lie within one arcminute ($\sim$0.5 Mpc) of one another, the size of a
group or small cluster of galaxies (see Waddington et~al.~1997 and
Pascarelle et~al.~1996 for the respective images).

In June 1997, we observed seven of the red cluster candidates using
the LDSS-2 multi-object spectrograph on the 4.2m William Herschel
Telescope.  A total of 40,000 seconds was obtained on each source, with
seeing of $\approx$1.0\arcsec.  Preliminary results from a rough
reduction of the first night of the data show that two of the sources
have strong emission or absorption features that put them at $z<1$.
The other five targets have definite continuum detections visible in
the two-dimensional images, but require careful sky subtraction before
the reality of the spectral features can be confirmed.  The inclusion
in the selected sample of two low redshift sources is entirely
consistent with the field contamination expected, and so far does not
invalidate our selection technique.

In conclusion, we have discovered ten sources with red $J-H$ colours
around the radio galaxy 53W002, that are consistent with the presence of a 
4000\AA-break at the cluster redshift.  The ages of the galaxies deduced from 
the strength of the break are consistent with a 1 Gyr burst model of galaxy 
evolution, observed at ages of 0.4--1.1 Gyr.  With the close 
proximity of these companions to one another and their relatively large 
size, the evidence suggests that 53W002 lies in a group or cluster that has 
already undergone significant evolution at $z=2.39$.

\acknowledgements We thank James Dunlop, John Peacock and Rogier 
Windhorst for their ongoing contributions to this work.  IW gratefully 
acknowledges support from the PPARC.

\end{document}